# Beam Cooling with ionisation losses.

C. Rubbia

INFN, Sezione di Pavia, Italy and CERN, Geneva, Switzerland

A.Ferrari, Y.Kadi and V. Vlachoudis

CERN AB Department, Geneva, Switzerland

Abstract.

This novel type of particle "cooling", called *Ionization Cooling*, is applicable to slow ($v \approx 0.1c$) ions stored in a small ring. The many traversals through a thin foil enhance the nuclear reaction probability, in a steady configuration in which ionisation losses are recovered at each turn by a RF-cavity. For a uniform target "foil", typically few hundred $\mu g/cm^2$ thick, transverse betatron oscillations are "cooled", while the longitudinal momentum spread diverges exponentially since — in the case of $dE/dx$ — faster (slower) particles ionise less (more) than the average. In order to "cool" also longitudinally, a chromaticity has to be introduced with a wedge shaped "foil", such as to instead increase (decrease) the ionisation losses for faster (slower) particles. Multiple scattering and straggling are then "cooled" in all three dimensions, with a method similar to the one of synchrotron cooling, but valid for low energy ions. Particles then stably circulate in the beam indefinitely, until they undergo for instance nuclear processes in the thin target foil.

This new method is under consideration for the nuclear production of a few MeV/A ion beams. Simple reactions — for instance $Li^7 + D \rightarrow Li^8 + p$ — are more favourably produced in the "mirror" kinematical frame, namely with a heavier ion colliding against a gas-jet $D_2$ target. Kinematics is generally very favourable, with angles in a narrow angular cone (around $\approx 10$ degrees for the mentioned reaction) and a relatively concentrated outgoing energy spectrum which allows an efficient collection as a neutral gas in a tiny volume with a technology at high temperatures perfected at ISOLDE.

It is however of a much more general applicability. The method appears capable of producing a "table top" storage ring with an accumulation rate in excess of $10^{14}$ Li-8 radioactive ion/s.

## 1.— . Introduction

Stochastic cooling and electron cooling are both and well established cooling techniques. They are however rather complicated and not suited to the requirements of specific applications related to very low energy, intense currents of ions, for which a much faster process must be envisaged. The unique features of the slow moving, highly ionising and massive ions suggest the development instead of a novel method based on the configuration of Figure 1, based on the non-Liouvillian nature of the dE/dx losses.

The basic configuration consists of (1) an appropriate (small) storage ring, (2) a thin target "foil" which induces energy losses and (3) an accelerating RF cavity. An initially injected ion beam — after being captured by ionisation stripping of the thin target into its highest ionisation state — is permanently stored in the ring. An accelerating cavity of an appropriate voltage and sufficient longitudinal amplitude replaces continuously the energy losses of the stored beam maintaining the equilibrium (orbit) configuration.

This method, which we shall call "ionisation cooling" closely resembles to the synchrotron damping of relativistic electrons — with the energy loss in the thin gas target substituting the function of the synchrotron light. The main feature of this method is that it produces an extremely fast cooling, compared to other traditional methods.

At typical energies of nuclear reactions (few MeV/nucleon) the associated ionisation losses are up to several $GeV/(g/cm^2)$. The shortness of the particle range makes nuclear interaction probability very small. A very large number of incoming particles is required in order to produce the chosen nuclear event. Storage of cooled beam particles is intended to increment considerably the efficiency of nuclear collisions with the help of a large number of traversals, made stable by the cooling process.

Already known since the fifties — in particular pointed out by Oreste Piccioni at the times of the MURA studies, but ultimately unsuccessful for relativistic proton beams — the energy loss by ionisation will be taken as equivalent to a continuous "resistive" drag force acting on the test particle:

$$F = -\frac{\delta U(E)}{T_S \beta_o c} \quad [1]$$

where $\delta U(E)/T_S$ is the energy loss in the target per unit of time, which must be energy dependent in order to produce longitudinal cooling, and $\beta_o c$ is the particle speed.

As well known, in the synchrotron light cooling, the ultimate level of the shrinking of the beam is determined by the quantum excitation due to the discrete nature of the emitted photons. Similar effects are expected in the ionisation cooling because of the transverse excitation due to multiple Coulomb scattering and longitudinal excitation due to Landau energy loss fluctuations. Because of the specific nature of the slow multiple charged ions, namely their high specific ionisation and large mass, these effects produce an acceptably low final beam size, as the resultant between the damping and the growth mechanisms. Indeed the lifetime of each stored particle is determined ultimately by the occurrence of a nuclear interaction in the thin target foil.

This general method will be specifically discussed for the reactions Li-7(d,p) Li-8 and Li-6(He3,n) B-8, which are chosen as reactions for radioactive beams production of the mirror unstable nuclei Li-8 and B-8 (Figure 2). It is however of a much more general applicability.

Particularly attractive in this context are the so-called beta-beams [1] in which neutrinos or antineutrinos are produced by the decay of radioactive and adequately short lived $\beta^+$ or $\beta^-$ emitters in a Lorentz boosted frame which generates the appropriate neutrino energy. An accelerator complex, beyond the purpose of the present paper, has to be designed in order to bring accelerated radioactive elements in an appropriate racetrack configuration, such as to orient the decay neutrino beam in the direction of the detector, hundreds of kilometers away.

This method may also be used in order to produce strong fluxes of radioactive beams for applications in hadron therapy [2]. Ion beams, typically protons or C-12 are used in order to produce a well concentrated amount of radiation deep in the body in order to remove cancerous parts by radiation damage. The use of B-8 or Li-8 instead of C-12 will further enhance the radiation dose at the terminal point of the track, since these radioactive nuclei decay immediately into two alpha particles (Figure 2). In addition the positron emission of B-8 can be used to visualise on-line the radiated tissues with the help of PET tomography.

Likewise it is possible to produce radioactive C-11 particles instead of C-12 in order to combine radiation damage with a PET scan in real time, in order to visualize the actual spatial distribution of the deposited radiation.

Many other applications in a number of different fields may also take profit of intense beams of radioactive ions.

## 2. – Exploiting the reverse kinematics.

The two body reactions here described are Li-7(d,p) Li-8 and Li-6(He3,n) B-8. In the region of few MeV deuteron and helium-3, the typical cross section for the reaction Li-7(d,p) Li-8 is about 100 mb with (a maximum at 200 mb), while for the reaction Li-6(He3,n) B-8 it is about a fraction 10 lower (Figure 3).

The recoiling final ions of an accelerated beam of D and He-3 impinging respectively on a Li-7 or a Li-8 target will be distributed in the laboratory over a large angular distribution and with very small kinetic energies, typically a few MeV (Figure 4). Most of them will necessarily come to rest inside the Li target, since for instance the B-8 range at 2 MeV is about 0.5 mg/cm². However the ionisation cooling of a D (or He-3) stored beam on a thin Lithium foil will be possible with a stable equilibrium between heating due to the foil traversal and cooling (Figure 1).

However, in order to reach a large intensity of radio-nuclides, the power dissipated by the beam will inevitably require a very thin but fast moving Li target arrangement, probably in the liquid form. The radio-nuclides will therefore travel away within a very substantial mass of moving target material and their prompt extraction will be extremely difficult. Therefore another alternative has been chosen, based on the kinematics of the "mirror" system, namely with a beam of Li-7 or Li-6 hitting a gaseous target either of D or of He-3.

The kinematics is very favourable, with emission angles of the final Li-8 and B-8 products concentrated in a narrow angular cone typically (Figures 5a and 5b) around about 10 and 12 degrees respectively, with a convenient and relatively concentrated outgoing energy spectrum. In the Figures, also the penetration power of the emitted radio-isotopes in a thin tantalum stopper is indicated. It appears appropriate to ensure the slow down in a few, small and sufficiently thin ring shaped foils (e.g. about 2.5 µm), such as to permit the efficient and prompt extraction of the neutral atoms in a time which is short compared with the decay lifetimes. The circulating beam is obviously not affected, because of the difference in angles with the reaction products. According to a technology already perfected at ISOLDE the resulting neutral gas then can be channelled to an ion source, where it is ionised again, prior to acceleration to high energies with more or less conventional methods.

The total cross sections for the two processes are shown in Figure 3. The relation from standard to reverse kinematics is linear over the energies of interest, with energy conversion factors $T_{Li-7} = 3.483 T_D$ for the reaction Li-7(d,p) Li-8 and $T_{Li-6} = 1.994 T_{He-3}$ for the reaction Li-6(He3,n) B-8.

For the energies of interest and Li-6 or Li-7 nuclei, the sums of the nuclear elastic and inelastic reaction cross sections, producing the ejection of the particle from the beam, are typically of the order $\sigma_{loss} \approx 10^{-24} \ cm^2$. The corresponding integrated target thickness for 1/e absorption in $D_2$ and He-3 are respectively ≈ 3.3 and ≈ 5 g/cm², to be compared with a typical foil thickness of a fraction of mg/cm². With a typical $D_2$ target thickness of 0.3 mg/cm² the nuclear 1/e lifetime is of about n ≈ 10⁴ turns.

The gas target can be a supersonic jet of gas at very high speed and therefore it can carry away the large power deposited by the beam. Li-7 and Li-8 can be produced using the same Z = 3 beam, starting alternatively with each isotopically separated species, depending on the type of neutrino emitted. Since the target is made of fast moving gas, an intense heating to substan-tial temperatures can be allowed. However while the use of Deuterium is entirely straightforward, He-3 is a very precious gas and only relatively small amounts of gas are likely to be realistic.

Additional phenomena producing significant particle losses may occur, like for instance capture of an atomic electron, changing the magnetic radius of the particle. An increased particle loss will shorten the 1/e lifetime of the stored beam: it may be compensated by a higher injected current for a given beam stored current[1].

Because of its shortness, many usual phenomena, like for instance intra-beam scattering and non-linear resonances and couplings play no relevant role.

### 3. — Ionisation of heavy ions.

The phenomenology, as already indicated, is exemplified by the one for Li-6 and Li-7. The mean ionisation energy losses of Li ion fully stripped ($z_{Li}$ = 3) after a path-length $t$ (g cm$^{-2}$) are represented by the well known Bethe-Bloch equation:

$$-\frac{dE}{dx} = 0.30058 \frac{m_e c^2}{\beta_o^2} z_{Li6}^2 \left[ \frac{1}{2} \log \frac{2 m_e c^2 \beta_o^2 \varepsilon_{max}}{I^2 (1-\beta_o^2)} - \beta_o^2 \right] t \qquad [2]$$

where $I$ is the mean excitation of the target medium ($D_2$). Both the density and the shell corrections, neglected in the formula, have been included in the numerical values of Table 1 where the specific ionisation loss $\Gamma$ per unit thickness $t$ is given as a function of the ion kinetic energy T.

Table 1. Some ionisation parameters for Li-6 and Li-7 in the energy interval of interest.

| Energy T, MeV | dE/dx MeV/(mg cm$^{-2}$) | | $\sqrt{\langle Z \rangle^2}$ | | $\delta E$ (keV) for loss of 0.3 MeV | |
|---|---|---|---|---|---|---|
| | Li-6 | Li-7 | Li-6 | Li-7 | Li-6 | Li-7 |
| 5 | 3.356 | 3.355 | 2.94 | 2.89 | 10.646 | 9.866 |
| 10 | 2.120 | 2.014 | 3.00 | 3.00 | 11.751 | 10.205 |
| 15 | 1.660 | 1.573 | 3.00 | 3.00 | 13.340 | 11.714 |
| 20 | 1.356 | 1.329 | 3.00 | 3.00 | 14.752 | 12.999 |
| 25 | 1.116 | 1.092 | 3.00 | 3.00 | 16.102 | 14.174 |
| 30 | 0.965 | 0.890 | 3.00 | 3.00 | 17.331 | 15.323 |
| 35 | 0.861 | 0.778 | 3.00 | 3.00 | 18.428 | 16.309 |

---

[1] The useful number of events per unit time is proportional to the product (stored current) x (foil thickness) and it does not depend on the beam lifetime as long as the injection current is adjusted accordingly.

Typically at T = 25 MeV, $\Gamma$ = 1.116 $GeV/(gcm^{-2})$ for Li-6 and $\Gamma$ = 1.092 $GeV/(gcm^{-2})$ for Li-7. Calculated ionisation losses are large and they are actually growing at lower energies more slowly than the naïve expectation, $1/T \approx 1/\beta_o^2$.

In Table 1 we have also shown the average ionisation charges $\langle Z \rangle$ of the Li-6 and Li-7 ions, as a function of the kinetic energy T. For energies T > 10 MeV, ions are quickly fully stripped already at the first traversal.

Accidental recombination by electron capture may occur during the storage, eventually leading to beam particle loss, because of its wrong charge at the exit of the foil. A semi-empirical formula (Schalchter et al.) is especially suited to estimate bound electron capture processes over a large energy range and projectile-target combination. The cross section for electron recombination $\sigma_{rec}$ is

$$\sigma_{rec} = 4.4 \times 10^{-23} [cm^2] Z_{proj}^{3.9} Z_{tar}^{4.2} (T/u)^{-4.8} \qquad [3]$$

In the case of Li-6 on He-3, $Z_{proj}$ = 3, $Z_{tar}$ = 2 and T = 25 MeV ($T/u$ = 4.16 MeV), the resulting cross section is $\sigma_{rec} = 6.21 \times 10^{-23}$ $cm^2$, rather large, about sixty times the estimated total inelastic nuclear cross section. The recombination cross section is a very fast falling function of the ion kinetic energy. For instance for T = 35 MeV, the loss rate is reduced to $\sigma_{rec} = 1.24 \times 10^{-23}$ $cm^2$.

We note however that the subsequent ionisation cross-section of a partially stripped ion is much larger, of the order of $\sigma_{ioniz} \approx 10^{-17}$ $cm^2$, since already after an average energy loss in the target of ≈ 2 keV, the ion is fully stripped again. Therefore already for energy losses in the foil as low as $\delta U \approx$ 300 keV, corresponding to a target thickness of 0.27 mg/cm² (T = 25 MeV), the electron captures surviving at the exit of the last layer of the target are reduced by over one hundred-fold, i.e. a 1/e loss rate of about $\sigma_{rec} = 0.6 \times 10^{-24}$ $cm^2$ (He-3 and T = 25 MeV), which makes it comparable to the expected nuclear loss. According to formula [3], the reaction Li-7(d,p) Li-8, has a cross section $\sigma_{rec}$ which is much smaller, since $Z_{tar}$ = 1.

The actual distribution of the energy loss, producing a longitudinal heating term in the ionisation cooling process, is characterised by small values of the largest allowed kinetic energy transfer, typically $\varepsilon_{max}$ = 7 keV for Li-6 at 25 MeV. Therefore there is no appreciable Landau tail and fluctuations are in an excellent approximation Gaussian distributions, with r.m.s. widths listed in Table 1. For a typical energy loss $\delta U \approx$ 300 keV, Li-6 at T = 25 MeV, corresponding to a foil thickness of 0.27 mg/cm², the resulting r.m.s. energy width is $\langle \delta T \rangle$ = 16.1 keV.

The mean square projected scattering of a particle of momentum p, speed $\beta_o c$ and charge e traversing the foil and producing heating in both transverse planes is given by the well known formula:

$$\langle \delta x'^2 \rangle = \left( \frac{13.6 \ MeV}{\beta_o cp} \right)^2 e^2 \frac{t}{\Lambda} \left( 1 + 0.038 \log \left( \frac{t}{\Lambda} \right) \right)^2 \qquad [4]$$

where $t/\Lambda$ is the foil thickness in units of the radiation length. The radiation length is $\Lambda = 126$ (122) g/cm$^2$ for D and $\Lambda = 67$ g/cm$^2$ for He-3. With $\delta U = 0.3$ MeV and T = 25 MeV the thickness is about $t = 0.27$ mg/cm$^2$ both for Li-6 on He3 and for Li-7 on D$_2$, corresponding to an r.m.s. scattering angle $\langle \delta x' \rangle = 0.29$ mrad and $\langle \delta x' \rangle = 0.20$ mrad respectively.

In order to take into account the behaviour of the beam propagation through many turns, the effects of the energy compensating RF cavity must be also taken into account. These effects are discussed in detail later on.

### 4. – Simple transverse compression.

Let us for one moment neglect Coulomb multiple scattering and $dE/dx$ straggling. The energy loss $\delta U$ in the foil (the differential symbol $\delta$ throughout the paper represents variations relative to a single turn) is compensated by the RF cavity, which adds however momentum only in the longitudinal direction, resulting in a small, proportional reduction of the particle transverse angle (both vertical and horizontal)

$$\delta x' = -x' \delta p/p = -x' \frac{1}{\beta_o pc} \delta U \qquad [5]$$

The emittance A (area/$\pi$) in either plane of displacement $x$ and angle $x'$ is given by the well known formula:

$$A = \gamma x^2 + 2\alpha x x' + \beta (x')^2 \qquad [6]$$

where $\gamma = 1/\beta$ and $\alpha = -\beta'/2$, which is assumed to be zero since the target is presumably in an apical point for the betatron function $\beta$ (note that the kinematical speed variable has been indicated with $\beta_o$). Therefore a simplified formula is:

$$A = \frac{1}{\beta} x^2 + \beta (x')^2 \qquad [6a]$$

The corresponding emittance change (the classic adiabatic damping) is calculated differentiating [.6a]:

$$A + \delta A = \frac{1}{\beta} x^2 + \beta (x' + \delta x')^2 = A + 2\beta x' \delta x' + \beta (\delta x')^2 \approx A + 2\beta x' \delta x' \qquad [7]$$

where the last term can be safely neglected since of second order in $dx'$. Inserting [5] in [7] we find $\delta A = -2\beta x'^2 \delta p/p$. The explicit, turn to turn variation of the square of the particle angle is $x'^2 = A/\beta \sin^2(\varphi + \varphi_o)$, where $\varphi$ is the betatron phase advance and $\varphi_o$ the initial phase. Averaged over many turns, far from resonances, $\langle \sin^2(\varphi + \varphi_o) \rangle \approx 1/2$ and therefore:

$$\frac{\langle \delta A \rangle}{A} = -\frac{\langle \delta p \rangle}{p} = -\frac{1}{\beta_o pc}\langle \delta U \rangle = -\frac{\langle \delta U \rangle}{2T} \qquad [8]$$

The kinetic energy T of the beam is such that $\beta_o pc = 2T$ and therefore a 1/e reduction in emittance is achieved when the foil has subtracted twice the beam kinetic energy, obviously compensated by the RF. Under the action of the continuous acceleration, the transverse emittances are therefore decreasing exponentially. If $P_S$ is the ring period, the usual exponential decay constant is given by $A(t) = A_o \exp(-\alpha_\varepsilon t)$ with:

$$\alpha_x = \alpha_y = \frac{1}{P_S}\frac{\delta U}{2T} \qquad [9]$$

For T = 25 MeV and $\langle \delta U \rangle$ = 0.3 MeV, the purely transverse cooling constant is about n = 166 turns. This is an extremely fast cooling, if one considers that the typical revolution frequency of the small storage ring is in the range 5÷10 Mc/s.

In these conditions, like in the similar case of the synchrotron radiation, the transverse emittance will converge to zero. In the case of ionisation cooling, a finite equilibrium emittance is due to the presence of the multiple Coulomb scattering.

The corresponding emittance growth is evaluated. The multiple scattering is introducing at each passage a change $\delta x'$ in the angle $x'$, leaving the impact co-ordinate $x$ unchanged. The corresponding increment of the emittance $A_C \to A_C + \delta A_C$ is:

$$\delta A_C = 2\beta x' \delta x' + \beta(\delta x')^2 \qquad [10]$$

After n traversals, the resulting averaged emittance growth per turn will be

$$\langle \delta A_C \rangle = \frac{1}{n}\sum_n \delta A_C \approx \beta \langle \delta x'^2 \rangle \qquad [11]$$

since the first term of [10] is averaging to zero. The mean square projected scattering of a particle traversing the foil is given by the value of $\langle \delta x'^2 \rangle$ and it is directly proportional to the foil energy loss $\langle \delta U \rangle$.

Under the combined effect of the adiabatic damping due to acceleration and the excitation due to multiple scattering, an equilibrium emittance is reached, which can be simply evaluated equating the damping rate [9] to the average growth rate [11]:

$$\beta \langle \delta x'^2 \rangle \leftrightarrow \frac{A_\infty}{2T}\langle \delta U \rangle \quad \text{from which} \quad A_\infty = 2T\beta \frac{\langle \delta x'^2 \rangle}{\langle \delta U \rangle} \qquad [12]$$

We remark that the equilibrium emittance does not depend on the specificity of the target foil, since both $\langle \delta x'^2 \rangle$ and $\langle \delta U \rangle$ are proportional to the thickness.

For the specific examples with T = 25 MeV, Li-6 on He-3 ($\langle\delta U\rangle$=300 keV and $\langle\delta x'\rangle$=0.29 mrad) and Li-7 on D$_2$ ($\langle\delta U\rangle$=300 keV and $\langle\delta x'\rangle$=0.20 mrad) we find for the equilibrium emittances $A_\infty$ = 2.04 x 10$^{-5}$ $\beta\ rad\ m$ and $A_\infty$ = 6.6 x 10$^{-6}$ $\beta\ rad\ m$ (area/π) respectively. Note that with the envisaged lattice structure, in view of the small circumference of the ring, the value of β at the target point is typically << 1 m and therefore acceptably small "accelerator like" equilibrium emittances are obtained.

### 5. – Longitudinal motion.

As well known, synchrotron radiation produces damping also in the longitudinal component, since the energy losses exhibit a significant momentum dependence ($\propto p^4$). A faster particle is loosing more energy and a slower particle less energy. In the case of the ionisation cooling, the opposite effect occurs, namely ionisation losses decrease with energy. This effect has to be studied in more detail.

The longitudinal motion of any particle is referred to its deviations from the synchronous particle. The momentum balance over one turn can be expressed as

$$\frac{d(\Delta p)}{dt} = \frac{1}{C_S}\left\{e\hat{V}[\cos(\varphi_S + \Delta\varphi) - \cos\varphi_S] - U(p)\right\} \qquad [13]$$

where $\Delta p = p - p_S$, $C_S$ is the circumference of the ring for the synchronous particle and the energy loss in the foil $U(T)$ per turn , is now a function of the particle kinetic energy T and momentum $p$.

A complementary first order differential equation [14] relates on how the changes in revolution frequency and equilibrium orbit, corresponding to the change $\Delta p$ affects $\Delta\varphi$:

$$\Delta p = \frac{E_S C_S}{2\pi h \eta}\frac{d(\Delta\varphi)}{dt} \qquad [14]$$

Combining [13] and 14] and in the small amplitude approximation, in order to simplify the cosine term, leads to a second order differential equation:

$$\frac{d^2(\Delta p)}{dt^2} + \frac{1}{C_S}\frac{dU}{dt} + \frac{2\pi h \eta e\hat{V}\sin\varphi_S}{E_S C_S^2}\Delta p = 0 \qquad [15]$$

The loss term can be taken as linear in a first approximation:

$$\frac{1}{C_S}\frac{dU}{dt} = \frac{1}{C_S}\frac{dU}{dp}\frac{d(\Delta p)}{dt} \qquad [16]$$

After some simple algebra Eq. [15] can be written as

$$\frac{d^2(\Delta p)}{dt^2} + \frac{1}{C_S}\frac{dU}{dp}\frac{d(\Delta p)}{dt} + \Omega_S^2\Delta p = 0 \quad ,\Omega_S = \left[\frac{2\pi h \eta}{C_S^2}\frac{e\hat{V}\sin\varphi_S}{E_S}\right]^{1/2} \qquad [17]$$

where $\Omega_S$ is the synchrotron frequency. It may be more convenient to use the energy deviations with the equation

$$\frac{d^2(\Delta E)}{dt^2} + \frac{1}{P_S}\frac{dU}{dE}\frac{d(\Delta E)}{dt} + \Omega_S^2 \Delta E = 0 \qquad [18]$$

where $P_S$ is the revolution period. This is the equation of a damped oscillator and its solution will be of the form

$$\Delta E(t) = A\exp(-\alpha_\varepsilon t)\cos(\Omega_S t - B) \quad , \quad \alpha_\varepsilon = \frac{1}{2P_S}\left[\frac{dU}{dE}\right] \qquad [19]$$

A similar procedure for $\Delta\varphi$ leads to an equation similar to Eq. [19]. Three different motions in $(\Delta\varphi, \Delta E)$ are possible, according to the actual value and sign of $\alpha_\varepsilon$ which is proportional to $dU/dE$:

- For $\alpha_\varepsilon > 0$, like it is the case for synchrotron radiation the energy loss of the foil per turn increases with the energy E and $(\Delta\varphi, \Delta E)$ phase space is an elliptical spiral; all particles of the beam are exponentially "cooled" as a function of time, moving toward the synchronous particle.
- For $\alpha_\varepsilon = 0$, namely an energy independent energy loss, the $(\Delta\varphi, \Delta E)$ motion is Liouvillian.
- For $\alpha_\varepsilon < 0$ — like it is the case for the foil — the energy loss per turn decreases with the energy E and the $(\Delta\varphi, \Delta E)$ phase space is growing exponentially (anti-damping).

As an example for the present case, we have set $U = 0.3$ MeV and a nominal (kinetic) energy T = 25 MeV. The derivative (see Table 1) in the case of Li-6 is $dU/dE = dU/dT = -0.0096$, corresponding to an e-fold increase of the initial longitudinal emittance after about 210 turns, i.e. the longitudinal momentum spread grows very fast, in contrast with the transverse cooling which has a e-fold constant of about n = 166 turns.

In order to introduce a change in the $dU/dE$ term — making it positive in order to achieve longitudinal cooling — the gas target may be located in a point of the lattice with a chromatic dispersion. The thickness of the foil must be wedge-shaped in order to introduce an appropriate energy loss change, proportionally to the displacement from the equilibrium orbit position.

An effective additional, linear variation of the energy lost by the foil as a function of the particle energy is introduced by a suitably shaped target, with thickness which is growing linearly with the radius of the orbit. In this situation, faster particles, (with a larger radius), will find a thicker wedge and consequently an increased ionization loss, such as to reverse the "natural" behaviour of a decreasing energy loss as a function of the energy.

As example we choose 25 MeV Li-6 ions and a wedge loss of 0.3 MeV for the synchronous momentum $p_s$. We show in Figure 6 four indicative examples of the resulting dependence of the ionization losses $\Delta U$ as a function of the momentum $\Delta p = p - p_S$: (1) the natural, unperturbed situation without the wedge effect, for which $dU/dE$ = - 0.0096, corresponding to a 1/e growth after n = 208 turns; (2) the wedge conditions in which $dU/dE \approx 0$, corresponding to no longitudinal cooling; (3) an increased wedge dependence in which $dU/dE$ = + 0.0094 and a 1/e cooling after n = 213; (4) the case of electrons of the same momentum (p = 550 MeV) and only natural synchrotron cooling (no wedge, natural losses $\propto p^4$) corresponding to $dU/dE$ = + 0.0243 and a 1/e cooling after n = 82 turns. It appears that already with a relatively moderated wedge one can realise conditions of the kind of synchrotron electrons.

The chromaticity of the lattice, namely the horizontal displacement due to momentum change at the point of the foil is defined as $\Delta x_p = D_x \Delta p/p$. In turns the wedge action will affect the energy loss proportionally to the displacement from the equilibrium orbit $x_S$, $U = U_o + \Delta x U' = U_o + D_x U' \Delta p/p$. The actual shape of the wedge depends on the chromaticity in the lattice, namely its shape and it is not defined for the moment since what matters in Figure 6 is $\Delta U = U - U_o = V \Delta p/p$, with $V = D_x U'$.

In the cases (1)-(3) respectively we find $V$ = 0 (flat wedge), $V$ = 0.48 MeV and $V$ = 0.95 MeV, namely $\Delta U/U_o$ =0.016 and $\Delta U/U_o$ = 0.0316 for each 1 percent of relative momentum change $\Delta p/p$ (cases (2) and (3)). Over the accessible momentum range therefore the impact of the wedge remains very modest.

The exponential decay of the longitudinal damping constant is counterbalanced by the Landau-Gaussian fluctuations in the energy losses, (in the synchrotron case it was due to quantum fluctuations) with an ultimate equilibrium between cooling and heating, corresponding to a given momentum spread.

For a typical energy loss $\delta U \approx$ 300 keV, Li-6 at T = 25 MeV, corresponding to a foil thickness of 0.27 mg/cm², the resulting r.m.s. energy width is $\langle \delta T \rangle$ = 16.1 keV. This term is better estimated with the numerical calculations which are presented later on.

### 6. – Excitation of horizontal betatron oscillations.

When traversing the absorber, the energy loss introduces also a displacement of the closed orbit and hence it introduces small changes in the transverse emittance. For a wedge absorber, its effect depends on the actual impact point and excitation of horizontal betatron oscillations is induced. Therefore a small orbit additional displacement is due to the change of the momentum with a corresponding change in the position of the equilibrium orbit

For a tiny change in longitudinal emittance after a single traversal, a change in position of the equilibrium orbit (x, x') and no change in direction $\delta x' = 0$ introduces an emittance change

$$A + \delta A = \frac{1}{\beta}(x + \delta x)^2 + \beta x'^2 = A + \frac{2}{\beta}x\delta x + \frac{1}{\beta}(\delta x)^2 \approx A + \frac{2}{\beta}x\delta x \qquad [20]$$

where the last term is neglected because of second order in $\delta x$ and $\delta x = D_x \delta p/p = D_x \delta T/2T_o$, where $D_x$ is the chromaticity of the lattice at the point of the wedge. The position of the orbit is related to the energy loss of the wedge $\delta T \equiv U = U_o + xU'$ at the orbit displacement x. Substituting we get for a single passage

$$\delta A = 2\frac{1}{\beta}x\delta x = \frac{D_x}{\beta T_o}(xU_o + x^2 U') \qquad [21]$$

Over many crossings the resulting time variation of A is then given as

$$\frac{dA}{dt} = \frac{\langle \delta A \rangle}{P_S} = \frac{1}{P_S}\frac{D_x}{\beta T_0}\left(\langle x \rangle U_o + \langle x^2 \rangle U'\right) \qquad [22]$$

where $P_S$ is the period of the ring, $x^2 = A\beta\cos^2(\varphi + \varphi_o)$, with $\varphi$ the betatron phase advance. Over many turns, far from resonance, $\langle \cos^2(\varphi + \varphi_o) \rangle \approx 1/2$ and $\langle x^2 \rangle = A\beta/2$ and $\langle x \rangle = 0$. Therefore in absence of a wedge action, namely $U' = 0$, the emittance is converging to zero and there is no transverse cooling contribution. More generally

$$\frac{1}{A}\frac{dA}{dt}\bigg|_{wedge} = \frac{1}{P_S}\frac{U'D_x}{2T_o} \qquad [22a]$$

The growth is counterbalanced by the damping due to the RF cavity, with a resulting horizontal damping rate

$$\frac{1}{A}\frac{dA}{dt} = \frac{1}{A}\frac{dA}{dt}\bigg|_{wedge} + \frac{1}{A}\frac{dA}{dt}\bigg|_{RF} = -\frac{1}{2T_o P_S}(U_o - U'D_x) \qquad [23]$$

from which we can have horizontal damping only provided $U_o > U'D_x$. Assuming for instance $2U_o = U'D_x$, we have the following damping rates:

$$\alpha_y = \frac{U_o}{2TP_S}$$
$$\alpha_x = \frac{1}{2TP_S}(U_o - U'D_x) \qquad [24]$$

The physical interpretation is therefore that the damping time for vertical betatron is the time needed to absorb by ionisation losses twice the kinetic energy of the particle, while the horizontal betatron is typically twice as long.

### 7. – Montecarlo simulation of the cooling process.

The whole process of betatron and longitudinal motion can be precisely described with the help of numerical simulation based on a Montecarlo model. Particle motion is followed at each turn both in the transverse planes $(x, x')$ and $(y, y')$ with the corresponding oscillation phase advance of the propagation matrix. No cross coupling is introduced in the transverse modes, although it could be easily introduced in the calculation.

Some of the main, indicative lattice parameters of the ring are given in Table 2.

Table 2.

| Lattice parameters | | | |
|---|---|---|---|
| Q-horizontal | $Q_x$ | 1.58 | |
| Q-vertical | $Q_y$ | 1.87 | |
| Gamma transition | $\gamma_t$ | 2.50 | |
| Orbit circumference | c | 4.00 | m |
| RF peak voltage | $V_o$ | 300.0 | kV |

The average values of the beta functions are correspondingly small, $\langle \beta_{x,y} \rangle = c/(2\pi Q_{x,y})$, because of the compact nature of the small ring. With circumference $c \approx 4$ m, the average value of the betatron function is then $\langle \beta \rangle \approx$ 0.35 m. At the position of the foil, betatron values are taken as $\beta_x^{Wedge} = \beta_y^{Wedge} =$ 0.15 m in both planes and the dispersion is $D_x = 0.15$ m. These parameters are indicative and they may be safely modified. The parameters of the gas target are $U_o = 300$ keV and it is wedge shaped with $U'x = 700$ keV/m.

The initial beam is chosen to be a fully ionised Li-7 at 27 MeV, corresponding to $\beta_o = 0.0980$ and $\gamma = 1.0048$. The period is $P_S = 0.136$ µs, corresponding to a revolution frequency $f_S = 7.35$ Mhz.

In correspondence of the RF cavity the lattice should be close to be achromatic i.e. with zero dispersion. If necessary, a bending magnet triplet with zero total bending could be inserted in order to enhance locally dispersion at the point of the wedge shaped gas target.

At each turn, traversing the gas target, the change in angle due to multiple Coulomb scattering and the Landau energy fluctuations of the energy loss are evaluated with the help of random generated Gaussians of appropriate r.m.s. width. This local change in the particle energy induces as well the appropriate betatron displacement of the equilibrium orbit. The wedge is inducing a linear function to the energy loss, $U_o + xU'$ according to the instantaneous position of the particle at point x.

In Figure 7 is shown a typical evolution of the first $10^4$ turns. Cooling is evidenced in both planes and in the momentum spread. The ratio of the equilibrium transverse emittances in the x-plane and y-plane are in the ratio 2:1, due to the presence of an appropriate x-slope in the foil thickness required to ensure longitudinal cooling. The fluctuations of the emittances are due to the finite number of particles generated.

## 8. – Achievable performance.

In absence of nuclear interactions, particles are circulating indefinitely. As already pointed out, for the energies of interest and Li-6 or Li-7 nuclei, the sum of the nuclear elastic and inelastic reaction cross sections, producing the ejection of the particle from the beam, are typically of the order $\sigma_{loss} \approx 10^{-24}$ $cm^2$. The corresponding integrated target thickness for 1/e absorption in $D_2$ and He-3 are respectively $\approx 3.3$ and $\approx 5$ g/cm², to be compared with a typical foil thickness of a fraction of mg/cm². With a typical $D_2$ target thickness of 0.33 mg/cm² the nuclear 1/e lifetime is of about n $\approx 10^4$ turns, namely the average beam particle duration in the ring is of the order of 1 ms.

The typical cross section for the specific reaction Li-7(d,p) Li-8 is about 100 mb, while for the reaction Li-6(He3,n) B-8 it is about a fraction 10 lower. Therefore the useful production yield is respectively about 10% and 1% of the incident Li beam yield. This is between three and four orders of magnitude greater than the one of a conventional thick target configuration with a single beam passage.

In practical unit and for an order of magnitude estimate, we may assume to produce $10^{14}$ reactions/s of Li-7(d,p) Li-8, corresponding to the beam injection of about $10^{15}$ ion/s. The injected current of singly ionised particles before the stripping is 160 μA. The corresponding injected beam power is relatively small, only 4 kW for T = 25 MeV.

The fully ionised beam in the ring is about a factor $10^{-3}$ smaller (nuclear beam lifetime $\approx 1$ ms), corresponding respectively to circulating intensities of $10^{12}$ for Li-7(d,p) Li-8, namely to circulating ion currents (Z = 3) of 3.5 Ampere for a revolution frequency $f_s$ = 7.35 Mhz. With an energy loss of 300 keV in the jet, the power due to the ionisation losses of the gas jet, generated by the re-circulating RF is 1.06 MWatt.

The radioactivity of such an intense source of the radioisotopes is very strong, of the order of 3400 Curies. The activity is mainly $\alpha$ and $\beta$ and from neutrons for the (not precisely known) alternate reaction Li-7(d,n) Be-8, which in turn decays instantly into 2 $\alpha$ and Li-7(d,2n) Be-7 with a cross section of about 100 mb. Be-7 has a half-life of 53 days and it decays $\beta^+$ back to Li-7 with a $\gamma$ line at 477 keV.

The emittances of Figure 7 are intended to show the ability of introducing a very substantial ionisation cooling. These correspond to extremely small beam sizes, since, for instance $A_y = 10^{-5}$ rad m and $\beta = 15$ cm corresponds to an r.m.s. vertical size of $\langle y \rangle = \sqrt{A_y \beta} = 1.2$ mm. Since the transverse heating is due to the Coulomb scattering, the cooled emittance is proportional to the local $\beta$ value. Higher $\beta$ values generate higher equilibrium emittance and therefore larger beam sizes, which in turn permit higher circulating currents.

The maximum circulating beam current, as equilibrium between incoming beam stripping and nuclear absorption, is primarily determined by the Laslett transverse tune shift due to space charges, given by the well known formula

$$N_{Laslett} = \Delta Q \frac{1}{r_p} \left(\frac{A}{Z^2}\right) 2(\pi\varepsilon)\beta_o^2 \gamma^3 B_F \qquad [25]$$

where $\Delta Q \approx 0.25 \div 0.5$ is the maximum tolerable Laslett tune-shift, $r_p = 1.53 \, 10^{-18}$ m is the classic proton radius, A = 7 and Z = 3 (for Li-7), $\pi\varepsilon = \pi(2\sigma)^2/\beta$ denotes the emittance area ($\varepsilon$ is the r.m.s. beam size), $B_F \approx 1/5$ is the bunching factor[2] and $\beta_o^2 \gamma^3 = 0.01$ are the usual relativistic parameters. Assuming for instance $\beta = 20$ m, the resulting emittance areas for the otherwise previously chosen parameters of the wedge have been calculated and are respectively as large as 2.4 x 10$^{-3}$ π rad m and 1.4 x 10$^{-3}$ π rad m, corresponding to a r.m.s. vertical size of 1.6 cm at the high beta point, which is reasonable.

The calculated number of circulating particles is then $N_{Laslett} = 10^{12}$ for the conservative estimate, $\Delta Q \approx 0.25$. If the energy loss in the foil can be made larger, the number of turns required to achieve nuclear absorption is correspondingly reduced. Since the storage time is shorter, the averaged number of particles produced is proportionally larger, at the price of a higher injection beam current.

While the required intensity of Li-7 can be obtained with one ring with the indicated parameters, even larger number of ions may be accumulated without exceeding a reasonable injected beam intensity. A possible way out is to share the incoming beam to a stack of 6 ÷ 8 rings, like it is the case for instance of the CERN-PS booster, having a common composite magnetic structure, a main RF source and a common structure for the gas jet, which of course should be able to dissipate a correspondingly larger cooling power. Since the structure of the storage ring is relatively simple, stacking of several rings should not constitute a serious problem and it may be easier than requiring a much larger beam cross section and a larger dissipated power in a single unit.

### 9. — The gas jet target.

The incoming beam is fully stripped to Z = 3 by the gas target and permanently stored in the ring. The gas target is also ensuring that cooling of the stored beam take place, as a necessary premise to nuclear collisions. We are referring to the previously mentioned parameters for Li-7(d,p) Li-8 reaction.

---

[2] Some additional improvements may be also possible decreasing the bunching factor B.

The gas jet target may follow the principle of a Supersonic Gas Injector (SGI) implemented for fuelling and diagnostics of high temperature fusion plasma [3] in several Tokamak, NSTX (USA), Tore Supra (France), HT-& and HL-1M (China), normally operated with $H_2$, $D_2$ and He gases. The technology is based on the isentropic compressible gas flow and it is well developed in aero-space, molecular beam research and industry. It is designed on a supersonic Laval nozzle with convergent and divergent contours (insert in Figure 9) which produces a highly uniform flow with Mach number M > 1, responsible for the formation of a low divergence high intensity jet. The stagnation volume with $P_o, T_o, \rho_o$ is followed by a narrow nozzle throat and a de Laval nozzle — at the end of which the exiting gas has $P < P_o, T < T_o, \rho < \rho_o$, and a Mach number $M_t > 1$ (Figure 9). The design is usually done using computational fluid dynamics based on the numerical solution of the Navier-Stokes equations.

Deuterium at pnt has a density of 1.8 x 10$^{-4}$ $g/cm^3$, corresponding to 5.4 x 10$^{19}$ $a/cm^3$. The gas jet target has an approximate thickness of 300 µg/cm² (9 x 10$^{19}$ atoms/cm²), equivalent to a target thickness of 1.6 cm of $D_2$ at pnt. The diameter of the jet may be of the order of 5 cm, providing the necessary gas thickness with a pressure $P \approx (1.6/5) \times 760$ $Torr$ = 250 Torr. Following ref. [3] on the SGI at NSTX and $D_2$, the jet velocity is about 2200 m/s, with a narrow divergence half-angle of ≈ 5 ÷ 12 degrees and $M_t \approx 4$. The volume of gas (at 250 Torr) is then 4.3 $m^3/s$, corresponding to $7.46 \times 10^{25} a/s$ or 248 $g/s$.

Our result is an extrapolation to larger sizes of the SGI of NSTX, and to scaling down of a large wind tunnel nozzle operated in air ($\gamma = 1.401$) at $P = 1$ atm and M = 8 [4]. In our case, the parameters are a nozzle throat diameter 3.36 mm and an inlet diameter of 29 mm for a prescribed exit diameter of 50 mm. The nozzle is 30.9 cm long. The pressure of the plenum at $M_t \approx 4$ is about $10 \times P[]$, namely $P_o = 2500$ Torr (3.3 atm). This is not critical and higher pressure with correspondingly higher $M_t$ are possible. Using isentropic relations, the temperature at the exit of the jet is about T = 70 K for $T_o = 300$ K.

The heat capacity of $D_2$ is 5200 J/(kg·K). The production of 1 MWatt of ionisation associated power corresponds to a temperature increase of 775 K through the passage of the jet. The outlet temperature is then about 570 °C. Such a relevant temperature change has to be carried away by the gas and cooled by an appropriate heat exchanger. The detailed numerical simulations have of course to include the effects of this sudden, large temperature heating of the gas.

In our case a residual pressure even of several Torr is acceptable, as long as the rest of the gas of the ring is "thin" when compared to the jet. The profile of the jet must be also capable of introducing an adequate wedge effect to ensure longitudinal cooling of the circulating beam. For this reason, instead of a single Laval geometry it is probably useful to replace it with an adequate number of smaller nozzles operating in parallel.

## 10. – Collection of produced ions.

As already pointed out, a nuclear reaction in the gas jet — like for instance Li-7(d,p) Li-8 — produces a secondary ion (Li-8) within a narrow angular spread (≈ 10 degrees) and with kinetic energy comparable with the one of the incoming heavy beam particle (see Figure 5). The range penetration of the secondary ion is very short, typically some tens of micron of solid material. The technique of using very thin targets in order to produce secondary neutral beams has been in use for many years. Probably the best known and most successful source of radioactive beams is ISOLDE [5].

A critical point is the determination of the delay of the collected ions to the catcher-ion-source system (CISS). This requires the understanding of the delay causing processes [6]. In essence the CISS as proposed here (Figure 8) is made of a small ring shaped thin box at about 10 degrees with a hole for the circulating ion beam and with a number of thin catcher foils inside. The release of the nuclide proceeds in two subsequent steps: *diffusion* from the place of implantation to the surface of the solid state catcher and *effusion* in the enclosure until the emission as a neutral particle.

Solid state diffusion is governed by Fick's law. For an initially homogeneous distribution with diffusion coefficient $D$ in a thin foil of thickness $d$, the release efficiency for an ion of half-life $\tau_{1/2}$ is

$$Y(\tau_{1/2}) = \tanh\left(\sqrt{\lambda \pi^2/4\mu_o}\right) / \sqrt{\lambda \pi^2/4\mu_o} \approx 0.76\sqrt{\mu_o \tau_{1/2}},$$

with $\mu_o = \pi^2 D/d^2$ and $\lambda = \ln 2/\tau_{1/2}$. Experimental information on the value of $D$ for Lithium in various (hot) materials is presently very contradictory and additional work is required before designing an appropriate CISS. It is however believed that at sufficiently high temperatures (≈ 2000 °C) an acceptable value is for $10^{-12} < D(m^2/s) < 10^{-10}$ as shown in Figure 10. The corresponding thickness is for the example of Tantalum is shown in Figure 11.

## 11. – Some consideration on the practical realisation of the accumulator.

On the basis of the previous considerations, a possible indicative set up is the following (Figure 8):

1) The incoming, stable and isotopically separated Li-7 may be conventionally accelerated for instance as a continuous beam by an appropriate accelerator, like for instance an RF-Q followed by a LINAC structure. Typical values are 25 MeV kinetic energy, singly ionised Li ions. The nominal current is about 160 µA, corresponding to a beam power of 4 kWatt. The beam frequency is likely to be different than the one of the accumulator and therefore the beam must be debunched and re-bunched at the frequency of the RF of the storage ring. No appreciable care is required both on the transverse and longitudinal emittances of the injected beam as long as it does not ex-ceed about 10$^{-3}$ π rad m. The beam transport is organised to ensure crossing with the circulating beam at 0° at the equilibrium position of the wedge.

2) The incoming beam is fully stripped to Z = 3 by the gas target and permanently stored in the ring. The supersonic gas jet target has an approximate thickness of 300 μg/cm² (9 x 10$^{19}$ atoms/cm²), an equivalent thickness of 1.66 cm at npt (D$_2$), corresponding to 5 cm at 250 Torr.

3) The gas jet is based on a supersonic Laval nozzle with convergent and divergent contours (insert in Figure 9) which produces a highly uniform flow with Mach number M > 1, responsible for the formation of a low divergence high intensity jet. Typically the jet velocity is about 2200 m/s, with a narrow divergence half-angle of ≈ 5 ÷ 12 degrees and a mach number $M_t \approx 4$. The pressure in the plenum is at least 3 – 4 atm at room temperature with a nozzle throat diameter of 3.36 mm. The volume of gas (at 250 Torr) is about 4.3 $m^3/s$, corresponding to $7.46 \times 10^{25} a/s$ or 248 $g/s$. Using isentropic relations, the temperature at the exit of the jet is about $T$ = 70 K for $T_o$ =300 K. The production of 1 MWatt of ionisation associated power corresponds to an outlet temperature is of about 570 °C. In our case a residual pressure away from the jet even of several Torr is acceptable, as long as the rest of the gas inside the ring is "thin" when compared to the jet. The profile of the jet must be also capable of introducing an adequate wedge effect to ensure longitudinal cooling of the circulating beam. The main power produced by the RF has to be dissipated by the gas jet and it is rather large, of the order of 1 MWatt, which has to be cooled by an appropriate heat exchanger and recompressed to the value and temperature of the plenum.

4) The detailed lattice of the ring is beyond the purpose of the present paper. It should be remarked however that the jet target must be in a high beta chromatic point ($\beta \approx$ 20 m, $D_X \geq$ 20 cm) while the RF cavity (-ies) should be located in an achromatic point.

5) The nominal RF voltage should be large enough to ensure the bucket containment during the full cooling cycle: a value V = 300 kV has been chosen. Note that Z = 3. The harmonic number should be low and the value h = 1 has been indicatively chosen. The cavity is operating at a fixed frequency (typically $f_S$ = 5 ÷ 10 Mhz).

6) The nuclear reactions in the gas jet — like for instance Li-7(d,p) Li-8 — in the "reverse" geometry produce a cone of secondary ions (Li-8) within a narrow angular spread and with kinetic energy comparable with the one of the incoming heavy beam particle (see Figure 5). The range penetration of the secondary ions is very short, typically some tens of micron of solid material. A possible method, following the best known and most successful source of radioactive beams of ISOLDE [5] consists in inserting some small and thin foils though which, according to Fink's law, the radioactive ions are diffused and collected for further use.

## 12. — Conclusions.

A novel type of particle "cooling", called *Ionization Cooling*, is discussed, based on slow ($v \approx 0.1c$) ions stored in a small ring. Singly ionised ions are fully stripped by a thin foil and remain permanently stored in the small ring. The ionisation losses due to many traversals through a thin foil compensated at each turn by an adequate RF-cavity. We observe that both multiple Coulomb scattering and straggling are "cooled" in all three dimensions, with a method similar to the one of synchrotron cooling, but valid for low energy ions. Particles circulate in the beam indefinitely, until they undergo for instance nuclear processes in the thin target foil. The nuclear reaction probability is greatly enhanced with respect to an ordinary target.

This new method is particularly suited for nuclear production of a few MeV/A beams. Simple reactions — for instance $Li^7 + D \rightarrow Li^8 + p$ — are more favourably produced in the "mirror" kinematical frame, namely with a heavier ion colliding against a gas-jet $D_2$ target. Kinematics is generally very favourable, with angles in a narrow angular and a relatively concentrated outgoing energy spectrum. This allows an efficient collection as a neutral gas in a tiny volume at high temperatures with the help of a technology perfected at ISOLDE.

## 13.— References

## 14.— Figure captions.

Figure 1.   Principle diagram of ionization cooling.

Figure 2.   The Li-Be-B isomeric triplet with A = 8.

Figure 3.   Total cross sections for the two indicated processes as a function of the equivalent D or He-3 energies and Li-6 and Li-7 at rest. The incoming energies for the "reversed" kinematics are also shown.

Figure 4.   Kinematics of the recoiling heavy ion for the reaction (1) $D + Li^7 \to p + Li^8$ with incoming D at 8 MeV and (2) $He^3 + Li^6 \to n + B^8$ with incoming He-3 of 15 MeV.

Figure 5.   Reverse kinematics for the two processes (a) Li-7(d,p) Li-8 and (b) Li-6(He3,n) B-8. The laboratory equivalent energies for Li at rest and the range in μ for Tantalum of the outgoing products are also shown. Marks correspond to 0.05 increments in the co-sine of the centre of mass angle.

Figure 6.   Energy loss/turn as a function of the momentum $\Delta p = p - p_s$ for cases 1-3 respectively of foil without wedge, and for 2 progressive increments of the wedge action. Case (4): synchrotron electrons.

Figure 7.   Computer simulation of the cooling of transverse emittances (area/π) and of relative longitudinal momentum spread as a function of number of turns. Cooling is evidenced in both planes and in the momentum spread. The ratio of the equilibrium emittances in the x-plane and y-plane are in the ratio 2:1, due to the presence of an appropriate x-slope in the foil thickness (see text), required to ensure longitudinal cooling. The value of the betatron function is taken to be 15 cm.

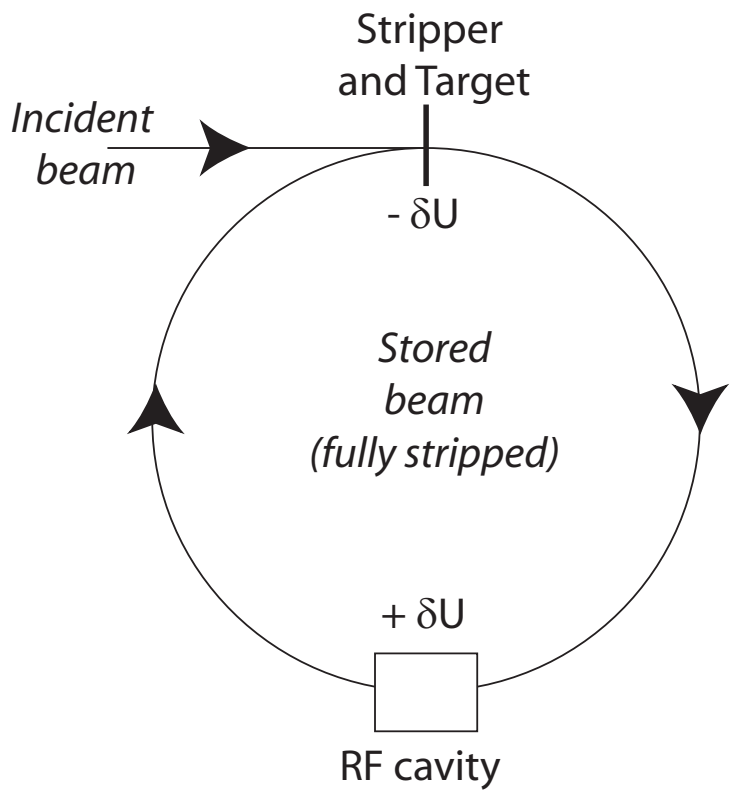

**FIGURE 1.**

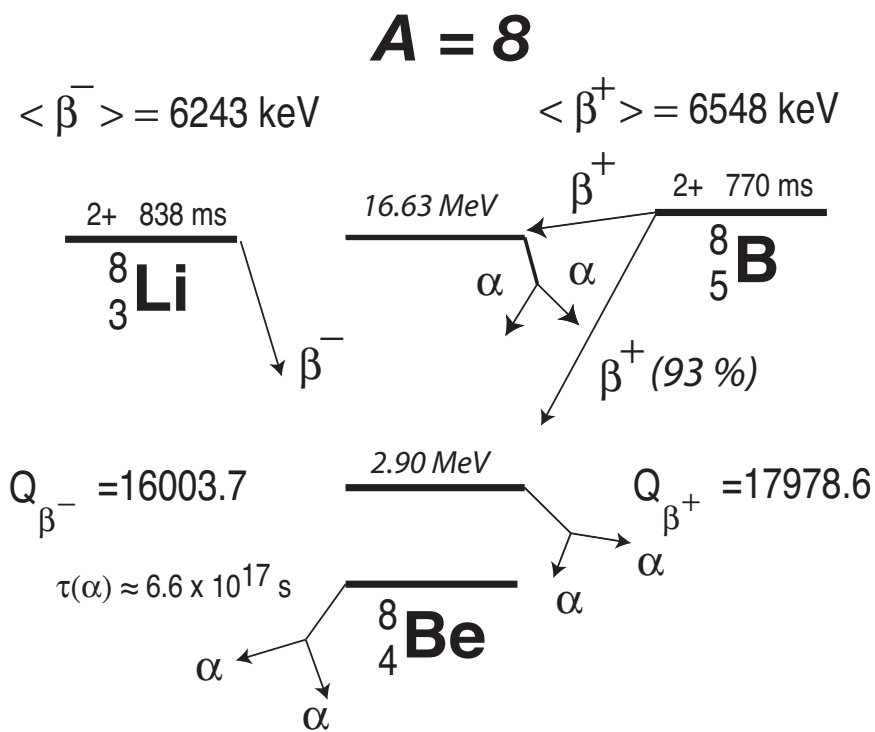

**FIGURE 2.**

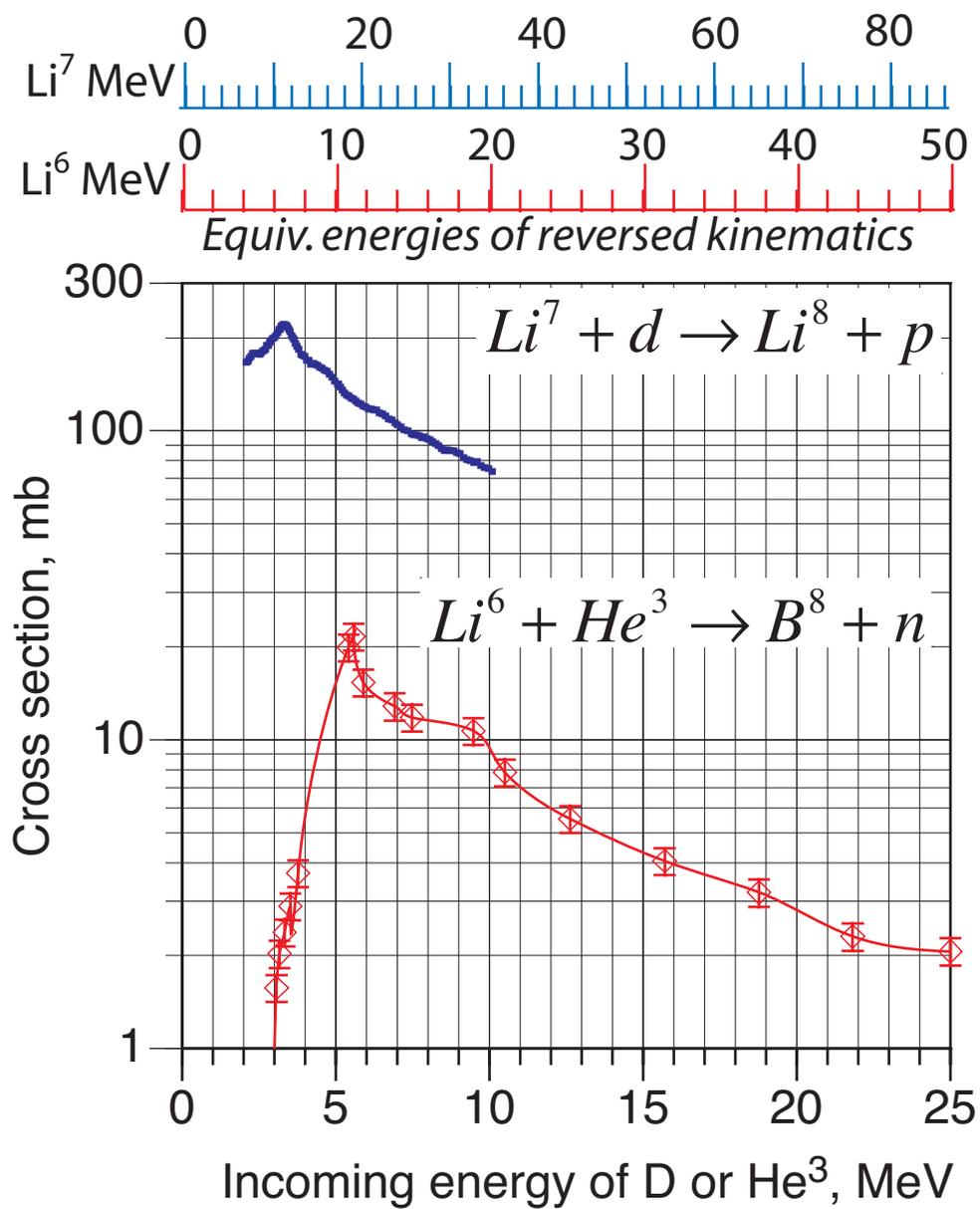

**FIGURE 3.**

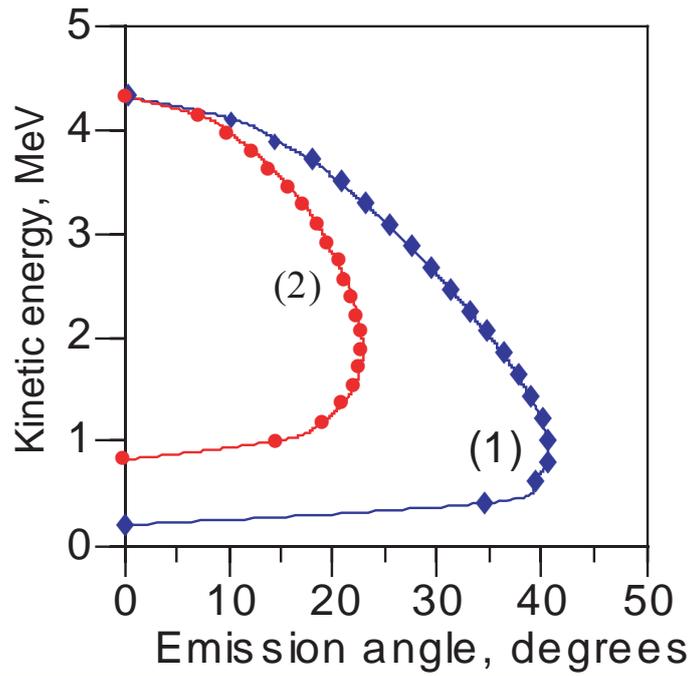

**FIGURE 4.**

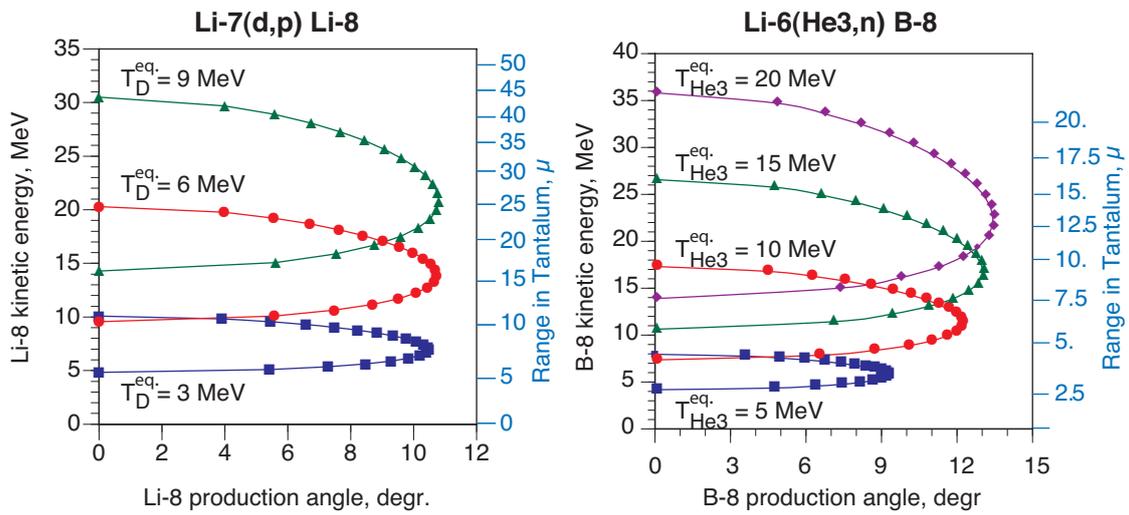

**FIGURE 5.**

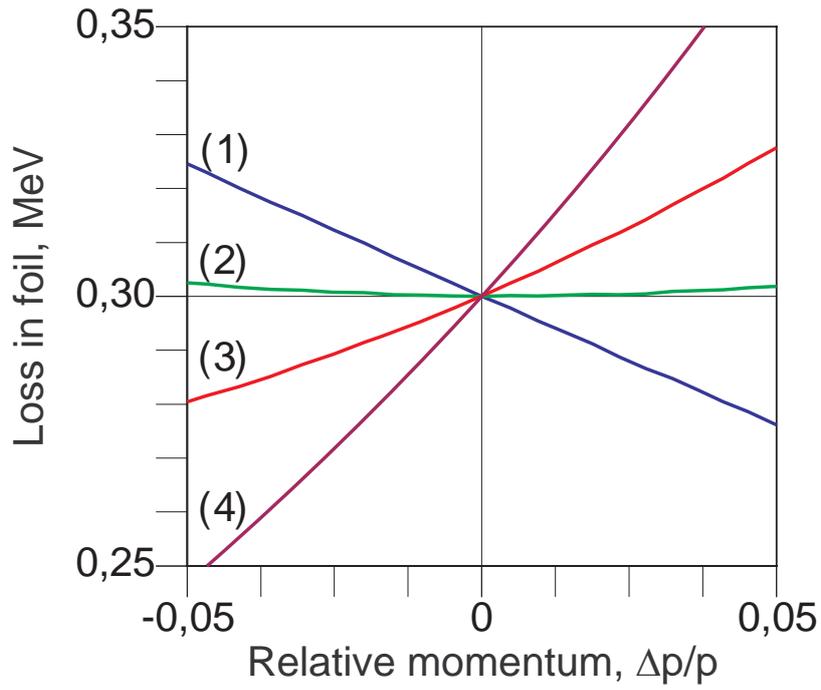

**FIGURE 6.**

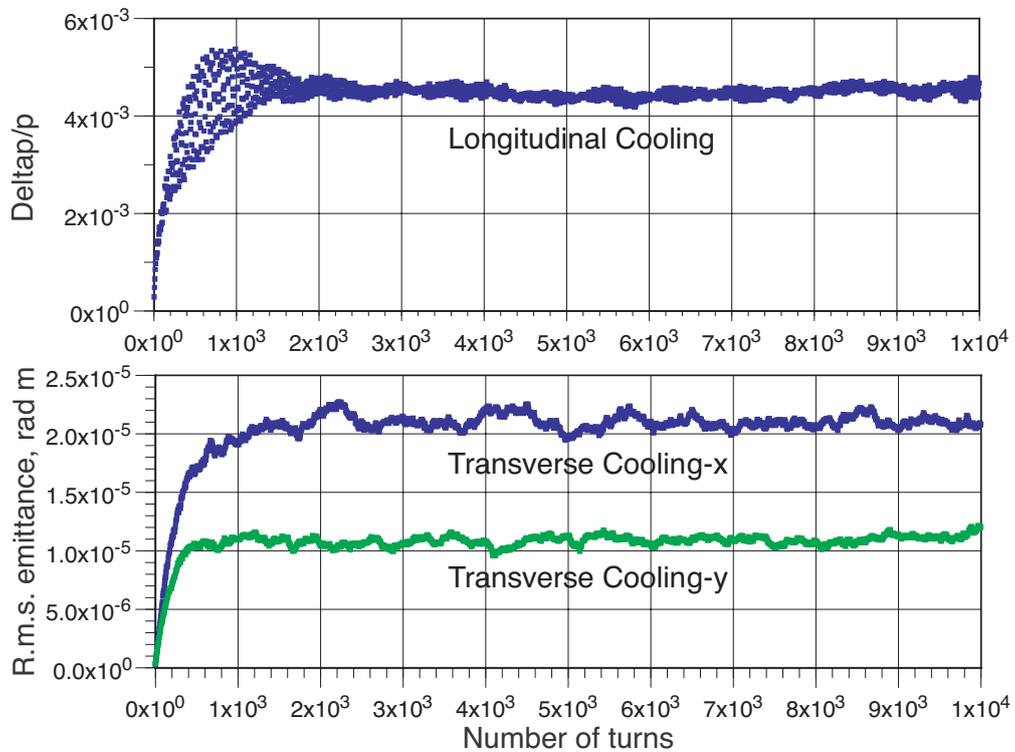

**FIGURE 7.**

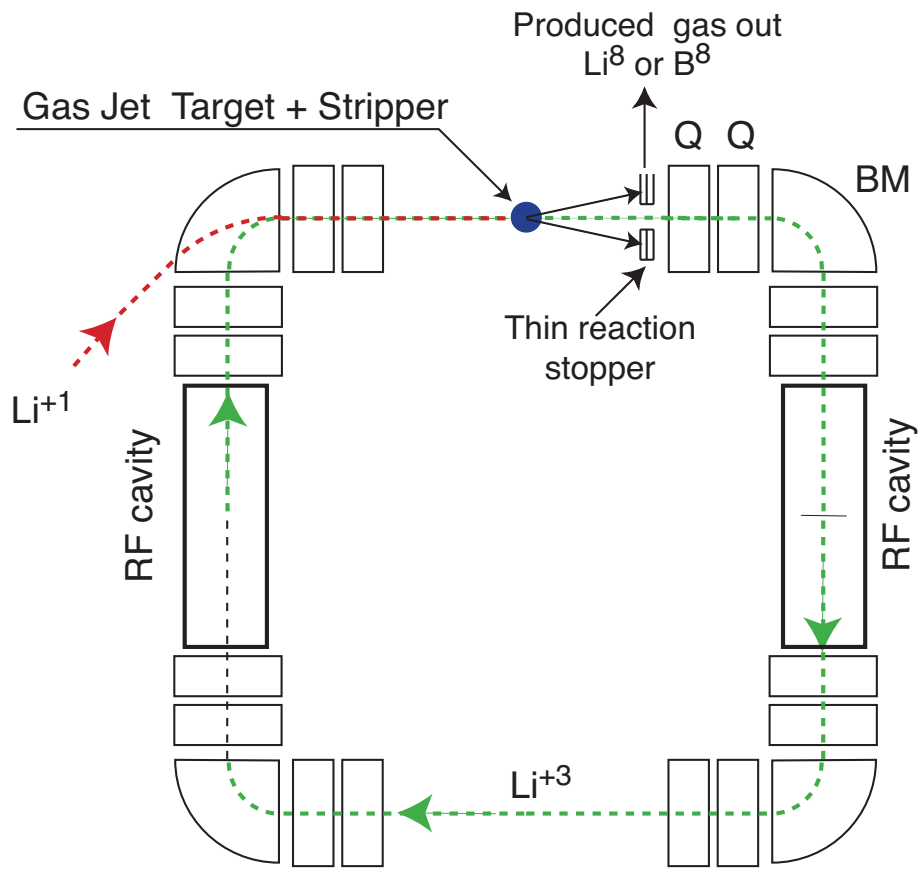

**FIGURE 8.**

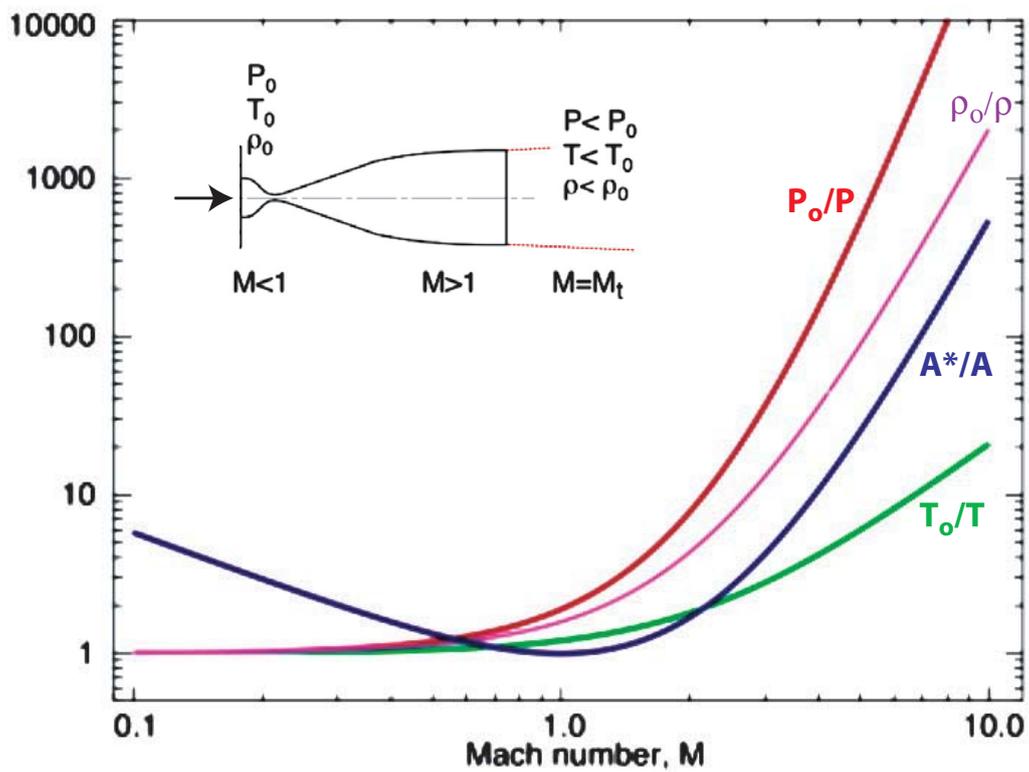

**FIGURE 9.**

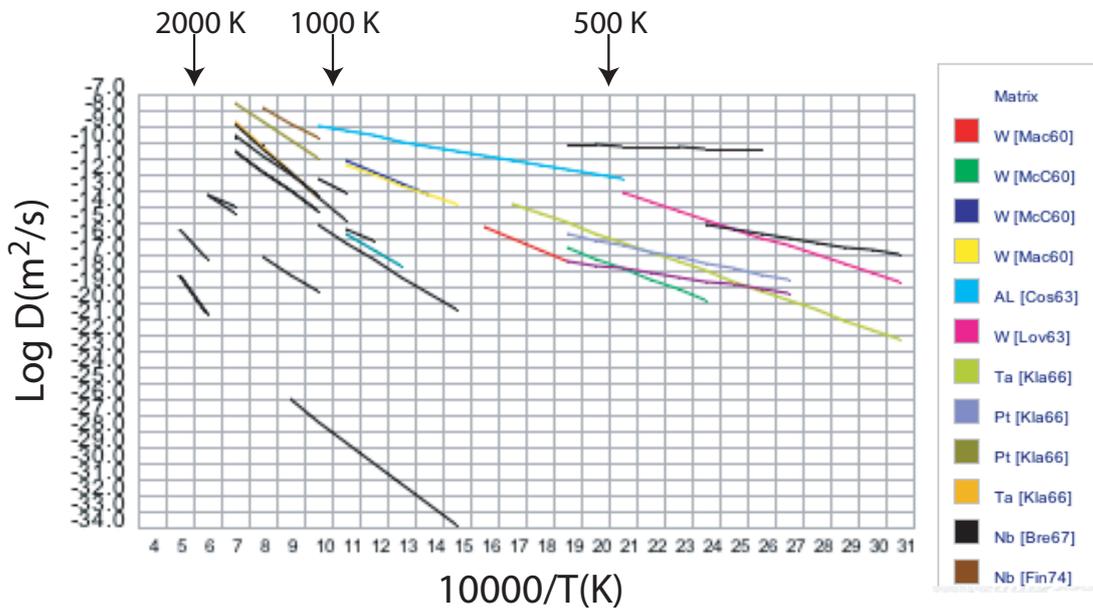

**FIGURE 10.**

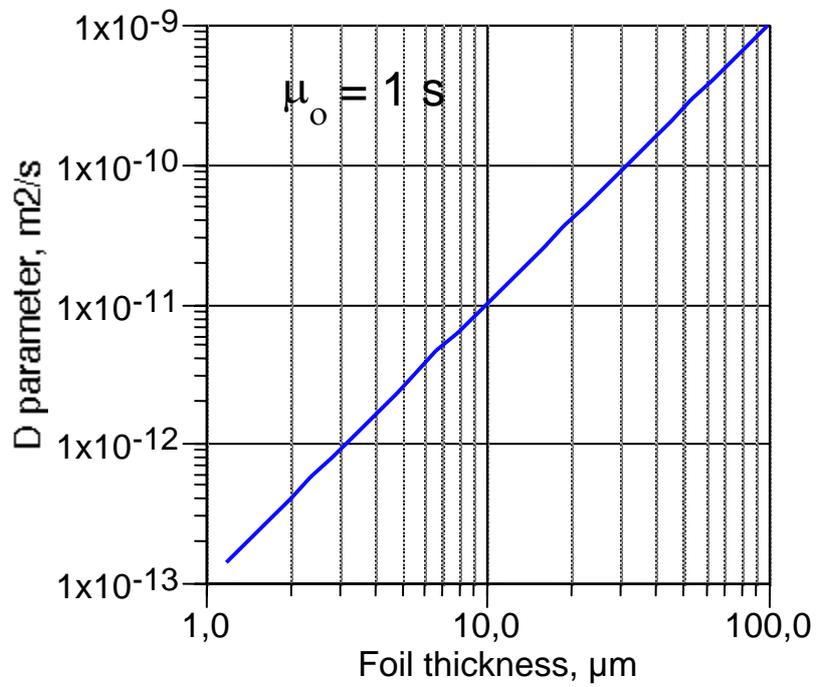

**FIGURE 11.**